\documentclass[10pt,a4paper,rorate,times]{article}
\usepackage[utf8]{inputenc}
\usepackage[T1]{fontenc}
\usepackage[affil-it]{authblk} 
\usepackage{multirow}
\usepackage{tabularx}
\usepackage{subcaption}
\usepackage[hidelinks]{hyperref}
\usepackage{url}
\usepackage{graphicx}
\usepackage{mathptmx}

\title{Social Network Datasets on Reddit Financial Discussion}
\author[1]{Zezhong Wang}
\author[2]{Siyang Hao}
\author[3,*]{Inez Maria Zwetsloot}
\author[4,5,6,*]{Simon Trimborn}
\affil[1]{Department of Systems Engineering, City University of Hong Kong, Hong Kong}
\affil[2]{College of Business, City University of Hong Kong, Hong Kong}
\affil[3]{Department of Business Analytics, Amsterdam School of Business, University of Amsterdam, The Netherlands}
\affil[1,4]{Amsterdam School of Economics, University of Amsterdam, The Netherlands}
\affil[5]{Tinbergen Institute, The Netherlands}
\affil[6]{Department of Management Sciences, City University of Hong Kong, Hong Kong}
\affil[*]{corresponding author(s): i.m.zwetsloot@uva.nl, simon.trimborn@uva.nl}
\date{} 

\begin{document}

\flushbottom
\maketitle

\begin{abstract}
	Stock markets are impacted by a large variety of factors including news and discussions among investors about investment opportunities. With the emergence of social media, new opportunities for having financial discussions arose. The market frenzy surrounding GameStop (GME) on the Reddit subreddit Wallstreetbets, caused financial discussion forums to receive widespread attention and it was established that Wallstreetbets played a leading role in the stock market movements of GME. Here, we present a new data set for exploring the effect of social media discussion forums on the stock market. The dataset consists of posts published on various Reddit subreddits concerning the popular meme stocks GameStop (GME), American Multi-Cinema Entertainment Holdings (AMC), and BlackBerry (BB). We document the data collection and processing steps and show that the posts and comments about these meme stocks are related to their market movements. 
\end{abstract}

\newpage

\section*{Background \& Summary}
Reddit is a social platform where millions of users globally communicate in different communities every day. A community, also called a subreddit, is organized according to users' interests, such as movies, music, programming, politics, and even finance. Users can post, vote, and comment in subreddits to exchange information and express opinions \cite{Reddit}. WallStreetbets (WSB), shown as r/wallstreetbets on Reddit, is a community where participants discuss stocks and share their investing strategies. It was even suggested that WSB contributes to financial literacy \cite{Klein2022}. WSB (\hyperlink{https://www.reddit.com/r/wallstreetbets/}{https://www.reddit.com/r/wallstreetbets/}) was created on January 31, 2012, and it has 15 million subscribers at the time of writing. It became notable for its major role in the GameStop short squeeze.

GameStop (NYSE: GME), an offline video game retailer, struggled with competition from digital distribution services and the negative economic impact of the COVID-19 pandemic. As a result, the stock price declined, and many hedge funds short-sold the stock because of the prediction of its further decrease. At that time, WSB users began to buy and hold GME stocks in bulk. Excess demand combined with limited availability drove the stock price and ultimately triggered a short squeeze. The stock price of GME rose from \$18.80 on 31 December 2020 to \$483 on 28 January 2021. Short sellers like Melvin Capital suffered considerable losses in this short squeeze. 

Sentiment can be a driver of market excesses and therefore various studies focused on the discussions on WSB and investigated the sentiment users reflect in posts and comments towards GME. It was indeed shown that sentiment on Reddit affects GME's stock price.\cite{long2022just, haq2022short, wang2021predicting, machavarapu2022reddit} Reddit, being a social network, allows also for the analysis of users interactions and the potential impact on others or even stocks. 
The foundation for network-based analysis is the existence of comment trees on Reddit. A comment tree consists of comments to a post and comments to comments within a post, which are the conversations between users. Various studies utilized users' interaction data on WSB to construct networks for analysis.\cite{zheng2021game, lucchini2022reddit, mancini2022self} These networks change over time because of the dynamics of users behaviour and are shown to relate to GameStop's stock price fluctuations. \textit{Lucchini et al. (2022)}\cite{lucchini2022reddit} identified a group of committed users who occupied central positions in the early conversation networks. These studies on sentiment and network interactions for GME and WSB have shown the existence of a relationship between WSB and GME's stock price. 

The articles mentioned above are based on different underlying data sets. \textit{Lucchini et al. (2022)}\cite{lucchini2022reddit} identified 36,128 events of commitment by 30,133 users on WSB from 27 Nov 2020 to 3 Feb 2021. \textit{Zheng et al. (2021)}\cite{zheng2021game} collected 6,330,271 posts and 8,666,673 comments published on the subreddit r/WSB from 08 Dec 2020 to 03 Feb 2021. \textit{Mancini et al. (2022)}\cite{mancini2022self} rely on 22,099,235 comments and 865,597 posts on WSB from 01 Sep 2019 to 01 Feb 2021. The Pushshift Reddit dataset, presented by \textit{Baumgartner et al. (2020)}\cite{baumgartner2020pushshift}, includes all the posts and comments published on Reddit between Jun 2005 and Apr 2019. In the meantime, access to the database is restricted to Reddit Moderators. A dataset focused on GME data across 4 subreddits was presented by \textit{J.J. Han (2022)}\cite{han2022reddit}. To allow for analysis of additional companies and the impact social networks such as Reddit have on them, we provide a dataset spanning more companies and subreddits to make this possible. Hence, in this data descriptor, we present three datasets for analyzing social networks' impact on stock markets. Besides GME data, our datasets include users' interactions focused on American Multi-Cinema (AMC) and BlackBerry (BB) on Reddit, which are three popular meme stocks. The data contain users, posts, interactions, and publishing time on Reddit, starting from 01.06.2020, 6 months before the first meme stock frenzy (GME) happened. For GME, the dataset ends on 31.08.2021, 8 months after the market frenzy took place, allowing for comprehensive post-event analysis. The market frenzies of AMC and BB took place later, hence their datasets end on the 31.12.2021. We design a framework for collecting and processing data from Reddit based on API and keywords, which can be generalized to other companies, topics of interest, or search queries. The datasets allow for several kinds of analysis, for example analyzing social networks' impact on the markets but also for network modeling, longitudinal network analysis, network monitoring, influencer identification, and bot detection. 

\section*{Methods}
\subsection*{Data collection}
\begin{figure}[ht]
	\centering
	\includegraphics[width=.9\linewidth]{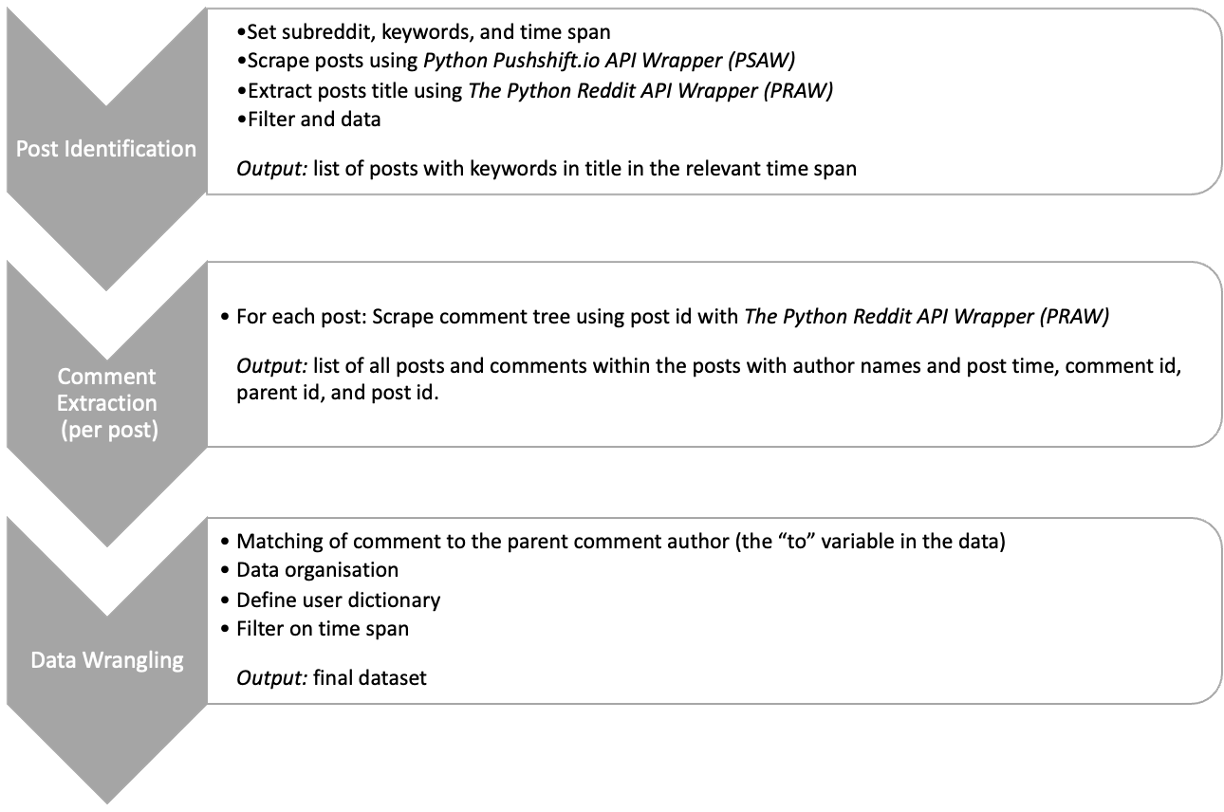}  
	\caption{Webscraping process step-by-step}
	\label{fig_webscraping}
\end{figure}

The data was obtained via the Reddit Data API. In this section, we describe in detail how the data was collected, processed and what actions were taken to validate the results and algorithms. The process consists of three steps, see Figure \ref{fig_webscraping}. Below we describe each step in detail in its respective subsection.

\subsection*{Post Identification}
The first step of the data collection process is `Post Identification'. This step results in a list of posts that fall within the search criteria. To obtain this list, one first needs to select the subreddit of interest, the keywords to search for, and the time period to include. Table \ref{Tab_post_indentification} shows an overview of the settings selected for our data collection efforts. For example, we collected GME-related data from the subreddit WallStreetBets from 01 Jun 2020 to 31 August 2021. We include all posts with the keywords `GME’ and/or `GameStop’ in their title. Note that we lowercase every character in the collected titles and compare them to ensure capitalization does not influence the results. We use the Python Pushshift.io API Wrapper (PSAW) \cite{PSAW} \hyperlink{Python Pushshift.io API Wrapper (PSAW)}{https://psaw.readthedocs.io/en/latest/} to obtain the post titles and ids.  This wrapper takes in a period, one keyword, and the subreddit of interest and returns a list of post ids. We repeat this procedure for each keyword and obtain multiple lists with post ids (one for each keyword). To ensure that the used wrappers work correctly, we have used PSAW to obtain post id's for the GME stock at two different time points, 3 months apart. We got the same results list with one post extra identified in the later validation. Then we use Python Reddit API Wrapper (PRAW) to extract the title associated with each post id. Finally, we need to clean and filter the obtained lists. This entails deleting duplicate posts (this may happen when there are multiple keywords in the title) and deleting those identified posts that have the keyword only in the main text (and not in the title).

\subsection*{Comment Extraction}
The second step of the data collection process is `Comment Extraction'. We use the Python Reddit API Wrapper (PRAW) \hyperlink{PRAW}{https://praw.readthedocs.io/en/stable/index.html} for extracting the comment tree post-by-post. Note that for the GME dataset, extracting the comment tree for eight posts gave us an HTTP413 response exception. For these eight posts we have therefore not included the comment trees in the final dataset. The result of this step is our datasets which requires cleaning and data wrangling.

\subsection*{Data Wrangling}
The final step is to perform data wrangling on the obtained list of posts and comments. The following actions are taken:
\begin{itemize}
	\item Matching of comments to parent comment author: for each comment, we identify the author of the parent comment and use this user id as the `to’ variable. 
	\item For the posts, we set the `to’ variable equal to the author’s id and the comment id equal to the post id.
	\item We define a unified user dictionary: each user name is recorded. 
	\item Filter on the time period to only include posts and comments in the period as mentioned in Table \ref{Tab_post_indentification}. Especially the comments collected can exceed these dates if the comment was posted (long) after the original post. 
	\item Set the timezone to UTC. 
\end{itemize}
This results in the final datasets.

\section*{Data Records}
\begin{figure}[ht]
	\begin{subfigure}{.5\textwidth}
		\centering
		\includegraphics[width=.9\linewidth]{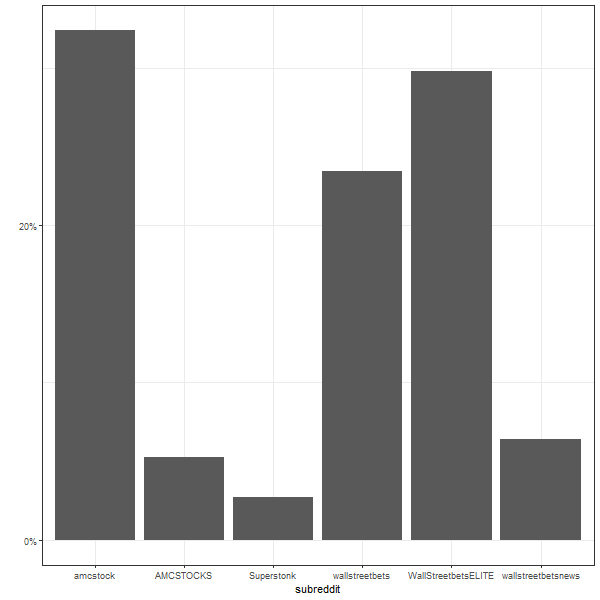}  
		\caption{Distribution of records from different subreddits for AMC}
		\label{fig_subreddit_AMC}
	\end{subfigure}
	\begin{subfigure}{.5\textwidth}
		\centering
		\includegraphics[width=.9\linewidth]{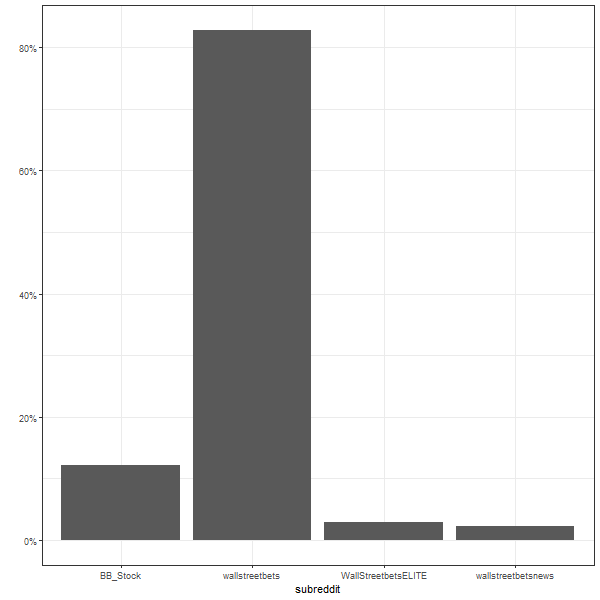}  
		\caption{Distribution of records from different subreddits for BB}
		\label{fig_subreddit_BB}
	\end{subfigure}
	\caption{Distribution of records from different subreddits for AMC and BB}
	\label{fig_subreddit}
\end{figure}

We provide a CSV file for the network data from Reddit for each stock. A user name dictionary for the three networks and a post dictionary are also available. Table \ref{Tab_files_summary} summarizes all files with brief introductions. The GME data are collected from the WSB subreddit as it was the most important subreddit for GME discussions during the frenzy. Besides WSB, the AMC data are collected from four additional subreddits, which were chosen due to the popularity of AMC in these forums. Likewise the BB data are collected from WSB and three additional subreddits. Figure \ref{fig_subreddit} shows the detailed subreddits of AMC and BB records. Comparative information about the three user interaction files is summarized in Table \ref{Tab_net_summary}.

Table \ref{Tab_var_summary} explains the variables in the interaction files. Each row indicates an interaction in the network, $from$ and $to$ indicate users on Reddit or vertices in the network. $post.id$, $comment.id$, and $parent.id$ determine the user behaviors and interactions, in other words, edges among vertices. These variables can describe the comment tree for a post on Reddit and build a network according to the following rules:
\begin{enumerate}
	\item When $from=to$ and $post.id=parent.id=comment.id$, the user $from$ published a post on Reddit. 
	\item When $from=to$ and $post.id=parent.id \neq comment.id$, the user $from$ comments to a post published by itself. 
	\item When $from=to$ and $post.id \neq parent.id \neq comment.id$, the user $from$ published a sub-comment to its comment in a post. 
	\item When $from\neq to$ and $post.id = parent.id \neq comment.id $, the user $from$ comments to user $to$'s post. 
	\item When $from\neq to$ and $post.id \neq parent.id \neq comment.id$, user $from$ published a sub-comment to user $to$'s comment in a post. 
\end{enumerate}

\begin{figure}[h]
	\begin{subfigure}{.45\textwidth}
		\centering
		\includegraphics[width=\linewidth]{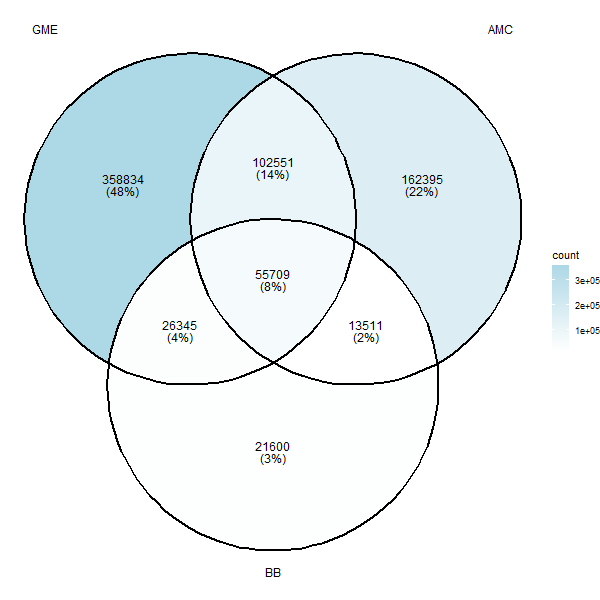}  
		\caption{Users' activities across different topics.}
		\label{fig_user}
	\end{subfigure}
	\begin{subfigure}{.45\textwidth}
		\centering
		\includegraphics[width=\linewidth]{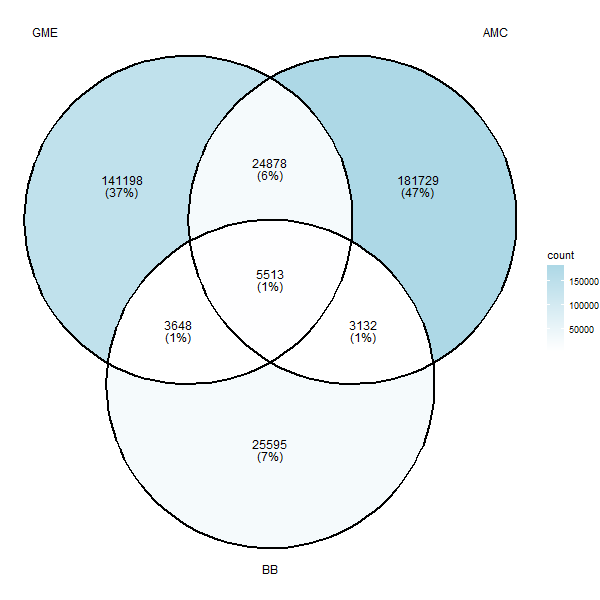}  
		\caption{Posts across different topics.}
		\label{fig_post}
	\end{subfigure}
	\caption{Users and posts by topic}
	\label{fig_user_posts}
\end{figure}

The `user{\_}dictionary.csv' file links the user to the three topics. 
This dictionary includes usernames and three bivariate variables representing users' participation in different topics. For example, users will have a label 1 for GME if they have participated in GME-related discussions; otherwise, they get 0. Figure \ref{fig_user} shows users' participation in different topics. 
Figure \ref{fig_post} shows number of posts in different topics.

\section*{Technical Validation}
In our datasets, the user name `None/NA' is abnormal, indicating a group of users who have deactivated or deleted their accounts or were shadowbanned by Reddit.
To eliminate the potential effect from these abnormal accounts, we remove `None'-related records before executing the following validation analysis.

\subsection*{Validation with reported events}
\begin{figure}[ht]
	\begin{subfigure}{.49\textwidth}
		\centering
		\includegraphics[width=\linewidth]{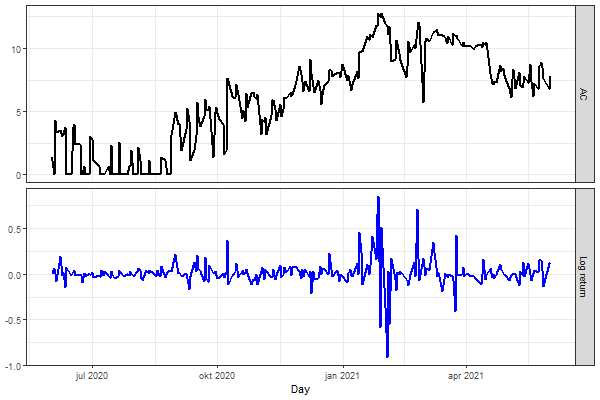}  
		\caption{Daily user activities regarding GME and its stock price}
		\label{fig_GME_trend}
	\end{subfigure}
    \begin{subfigure}{.49\textwidth}
    	\centering
    	\includegraphics[width=\linewidth]{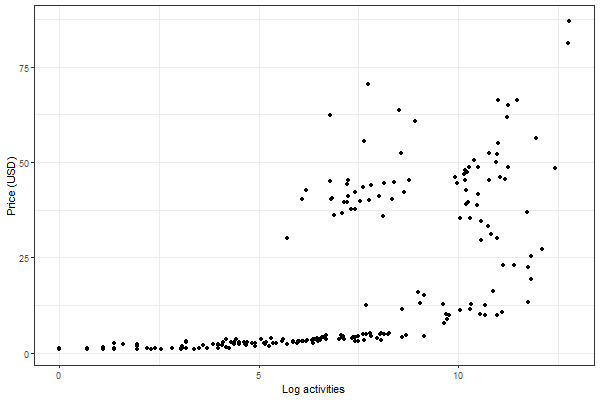}  
    	\caption{Scatter plot of GME daily user activities and stock price}
    	\label{fig_GME_scatter}
    \end{subfigure}
\\
	\begin{subfigure}{.49\textwidth}
		\centering
		\includegraphics[width=\linewidth]{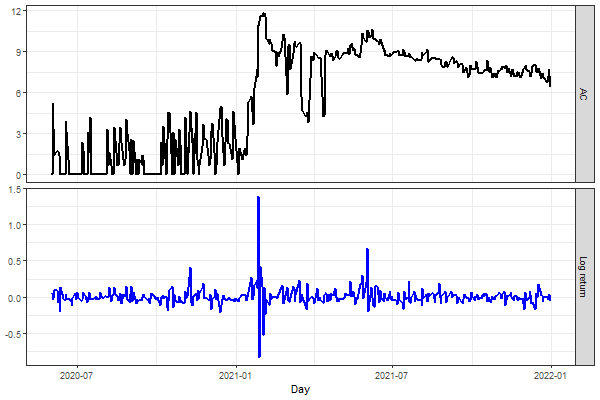}  
		\caption{Daily user activities regarding AMC and its stock price}
		\label{fig_AMC_trend}
	\end{subfigure}
	\begin{subfigure}{.49\textwidth}
		\centering	\includegraphics[width=\linewidth]{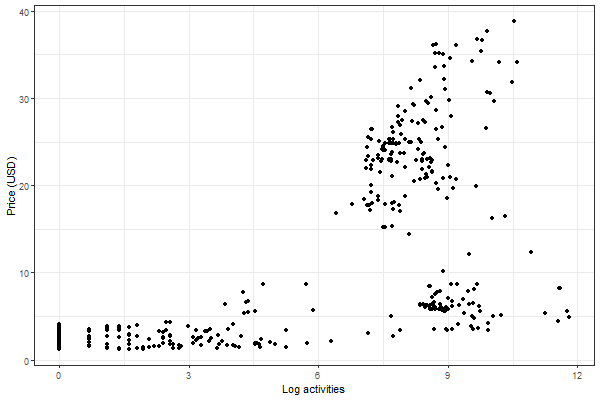}  
		\caption{Scatter plot of AMC daily user activities and stock price}
		\label{fig_AMC_scatter}
	\end{subfigure}
\\
	\begin{subfigure}{.49\textwidth}
		\centering
		\includegraphics[width=\linewidth]{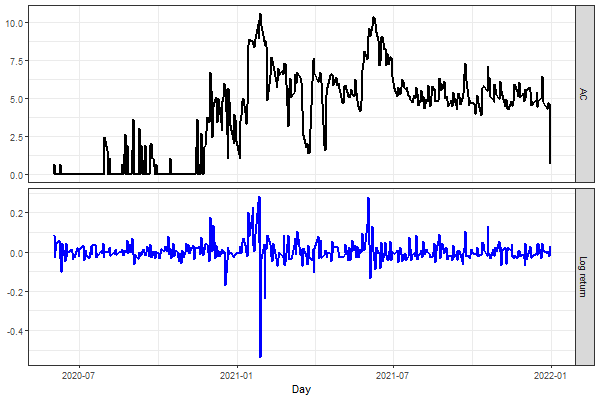}  
		\caption{Daily user activities regarding BB and its stock price}
		\label{fig_BB_trend}
	\end{subfigure}
	\begin{subfigure}{.49\textwidth}
		\centering
		\includegraphics[width=\linewidth]{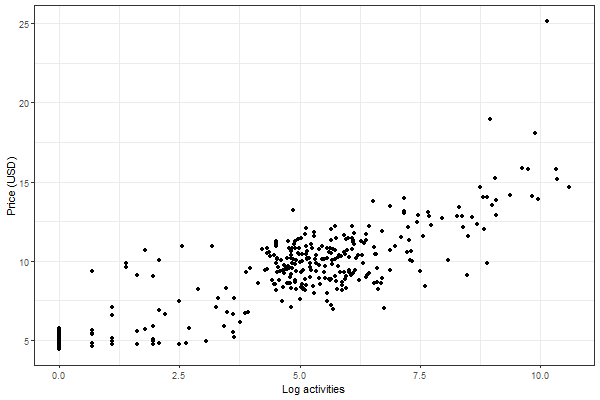}  
		\caption{Scatter plot of BB daily user activities and stock price}
		\label{fig_BB_scatter}
	\end{subfigure}
	\caption{Change of user activities and stock price}
	\label{fig_Validation}
\end{figure}

The literature review in Background \& Summary shows that users behavior on WSB can affect the stock price of GME, we use this conclusion to validate our data. We compare daily user activities about each topic with the corresponding daily stock price. The daily price data are easily accessed online, so we do not include and provide them in our data sets. We compute the logarithm of daily activities to make them scale comparable to daily stock price and draw the time series and scatter plots in Figure \ref{fig_Validation}. As shown in Figure \ref{fig_GME_trend}, user activities in WSB increase with the stock price, and both variables peaked in week 25 Jan 2021 to 31 Jan 2021; the GME short squeeze also occurred in this week. The scatter plot in Figure \ref{fig_GME_scatter} shows a positive correlation between stock price and Reddit users’ behaviors. 

Figure \ref{fig_AMC_trend} shows that the discussion about AMC on Reddit increased abruptly in Jan 2021, and the stock price increased in the same period. At the beginning of June 2021, user activities reached another peak, and the stock price soared from around $10$ to as much as $40$. Traders on the wrong side of that bet are exposed to severe losses. To the best of our knowledge, no research has analyzed the relationship between this market frenzy and Reddit users, which could be an interesting project. The scatter plot for AMC (Figure \ref{fig_AMC_scatter}) also shows that more user activities on Reddit relate to higher stock prices. Figure \ref{fig_BB_trend} shows two peaks in Reddit user activities and two corresponding peaks in stock price. Figure \ref{fig_BB_scatter} shows a strong positive correlation, which confirms the relationship between user activity and stock price. These two datasets can be used for factor analysis or predicting the stock price of AMC and BB.  

\subsection*{User behavior}
Figure \ref{fig_Validation} shows the temporal user activities on Reddit. As different users may have different characteristics, we count the number of activities for each user based on the three interaction files. Table \ref{Tab_user_summary} presents the summary statistics. The range of user activities is from 1 to thousands, and the standard deviations also show that users have quite different activity levels. Most users have few actions within the selected period about these three stocks; the medians are close to or equal to the minimum.

\section*{Usage Notes}
Our datasets are available at request for research purposes. As our datasets keep the detailed user interactions on Reddit, they can be immediately applied to a wide range of different research topics, including:

\begin{itemize}
	\item Relationship between user behaviors, stock prices, and big events. Research has shown that Reddit user behavior can affect GME stock prices and cause a short squeeze. Our datasets allow us to expand this conclusion to stocks such as AMC and BB. Also, user activities on Reddit may be impacted by other factors, such as policies, the economy, and company news. These factors could be used to explain changes in Reddit users behavior and then explain the changes in stock prices. Hence, involving extra events in the relationship analysis between Reddit and the stock market might be interesting. 
	\item Sentiment and corresponding structures. Post' titles have been frequently used for sentiment analysis\cite{anand2022role, long2022just, haq2022short}, hence we provide a post dictionary for the datasets. Identifying the relationship between sentiment in the title and the post structure can reflect which type of post title can attract more attention and participation from users, which can help Reddit to improve user experience and engagement. Combining post structure with sentiment analysis may also shed light on how Reddit affects the stock market. 
	\item Bot detection. Bot detection for social media is a hot topic since they have been extensively used for disingenuous purposes, such as perpetuating scams, swaying political opinions, and manipulating the stock market \cite{kudugunta2018deep}. We identify bots by the activity frequencies in data validation. There is a graph-based bot detection method in the political subreddit proposed by \textit{Hurtado et al. (2019)}\cite{hurtado2019bot}. We attempted to apply the graph-based method to the GME dataset, but the network is too big and complex to process the method in reasonable computing time. There are limited methods designed for identifying bots on Reddit. Hence, developing a bot detection algorithm for large scale social networks such as Reddit remains an open research question which can be tackled based on our dataset. 
	\item Network analysis. The provided interaction files can be used for network analysis, such as building a sparse network model, monitoring temporal changes, identifying influential nodes/users, and detecting changepoint. It was shown that network changes can be used to explain the return movements and realized volatility of GME\cite{network2023}.
\end{itemize}


\section*{Figures \& Tables}

\begin{table}[h]
	\begin{center}
		\captionsetup{justification=centering}
		\caption{Overview of Post Identification criteria}
		\begin{tabular}{m{8em} | m{16em}  m{7em}  m{11em}} 
			\hline \hline &\\[-2ex]
			Stock & Subreddit & Keywords & Time period \\[0.1ex]
			\hline &\\[-2ex]
			GameStop (GME) & wallstreetbets & GME, GameStop & 01 Jun 2020 - 31 Aug 2021 \\[2ex]
			AMC Entertainment (AMC) & wallstreetbets, WallStreetbesELITE, amcstock, AMCSTOCKS, Superstonk, wallstreetbetsnews & AMC & 01 Jun 2020-31 Dec 2021 \\[4ex]
			BlackBerry (BB) & wallStreetbets, BB\textunderscore Stock, WallStreetbetsELITE, wallstreetbetsnew & BB, BlackBerry, Black Berry & 01 Jun 2020 - 31 Dec 2021 \\[2ex]
			\hline \hline
		\end{tabular}\label{Tab_post_indentification}
	\end{center}
\end{table}

\begin{table}[h]
	\begin{center}
		\captionsetup{justification=centering}
		\caption{Summary of files}
		\begin{tabular}{ m{10em} | m{35em} } 
			\hline\hline &\\[-2ex]
			File name & Description \\[0.1ex]
			\hline &\\[-2ex]
			gme{\_}interactions.csv &  User interactions about GME from 01 June 2020 to 31 Aug 2021 UTC\\[2ex]
			amc{\_}interactions.csv &  User interactions about AMC from 01 June 2020 to 31 Dec 2021 UTC\\[2ex]
			bb{\_}interactions.csv & User interactions about BB from 01 June 2020 to 31 Dec 2021 UTC\\[2ex]
			user{\_}dictionary.csv & Include user names, user IDs, and stock\\[2ex]
			post{\_}dictionary & Include post title, post ID, posting time, posting time in utc, user{\_}no, and stock\\[0.5ex]		
			\hline\hline
		\end{tabular}\label{Tab_files_summary}
	\end{center}
\end{table}

\begin{table}[h]
	\begin{center}
		\captionsetup{justification=centering}
		\caption{Summary of network data for GME, AMC, and BB.}
		\begin{tabular}{m{9em} | m{11em} m{11em} m{11em}} 
			\hline \hline &\\[-2ex]
			Information & GME & AMC & BB\\[0.1ex]
			\hline &\\[-2ex]
			Variables  & from, to, time, utc, post.id, comment.id, parent.id, subreddit & from, to, time, utc, post.id, comment.id, parent.id, subreddit & from, to, time, utc, post.id, comment.id, parent.id, subreddit \\[4ex]
			Start Date & 2020-06-01 & 2020-06-01 & 2020-06-01 \\[1ex]
			End Date & 2021-08-31 & 2021-12-31 & 2021-12-31\\[1ex]
			Number of Records & 4,766,814 & 2,671,047 & 506,460\\[1ex]
			Number of Posts &  175,237 & 215,151 & 37,888\\[1ex]
			Number of Comments & 4,591,577& 2,455,795 & 468,572\\[1ex]
			Number of Users & 543,439 & 334,166 & 117,165\\[0.5ex]
			\hline \hline
		\end{tabular}\label{Tab_net_summary}
	\end{center}
\end{table}

\begin{table}[h]
	\begin{center}
		\captionsetup{justification=centering}
		\caption{Explanation of variables in interaction data files.}
		\begin{tabular}{ m{10em} | m{35em} } 
			\hline\hline &\\[-2ex]
			Variable & Description \\[0.1ex]
			\hline &\\[-2ex]
			from & A unique username for a Reddit user who publishes a comment or a post.\\[1ex]
			to & A unique username for a Reddit user who receives a comment or publishes a post.\\[1ex]
			time & The publishing time of comments and posts from user from.\\[1ex]
			utc & UNIX time stamp.\\[1ex]
			post.id & A string of characters with numbers and letters, unique for each post.\\[1ex]	
			comment.id & A string of characters with numbers and letters, unique for a comment. \\[1ex]
			parent.id & A string of characters with numbers and letters, unique for a comment or a post that the comment is replying to.\\[2ex]
			subreddit & The subreddit for corresponding activities.\\[1ex]
			\hline\hline
		\end{tabular}\label{Tab_var_summary}
	\end{center}
\end{table}

\begin{table}[h]
	\begin{center}
		\captionsetup{justification=centering}
		\caption{Degree (number of activities, both in and out degree) for users in GME, AMC, and BB.}
		\begin{tabular}{m{9em} | m{7em} m{7em} m{7em}} 
			\hline\hline &\\[-2ex]
			Statistics & GME & AMC & BB \\[0.1ex]
			\hline &\\[-2ex]
			Minimum & 1  & 1 & 1 \\[1ex]
			$10^{th}$ Percentile & 1 & 1 & 1\\[1ex]
			$25^{th}$ Percentile & 1 & 1 & 1 \\[1ex]
			$50^{th}$ Percentile & 2 & 2 & 2 \\[1ex]
			$75^{th}$ Percentile & 7 & 7 & 5 \\[1ex]
			$90^{th}$ Percentile & 22 & 21 & 12 \\[1ex]
			Maximum & 955,971 &75,701 & 37,481 \\[1ex]
			Mean & 15.9 & 14.4 & 7.7\\[1ex]
			Standard Deviation & 1322.2 & 224.5 & 140.0\\[0.1ex]
			\hline\hline
		\end{tabular}\label{Tab_user_summary}
	\end{center}
\end{table}

\end{document}